\documentclass[smallabstract,smallcaptions]{dccpaper}

\usepackage{epsfig}
\usepackage{citesort}
\usepackage{amsmath}
\usepackage{amssymb}
\usepackage{color}
\usepackage{url}

\newlength{\figurewidth}
\newlength{\smallfigurewidth}

\usepackage{amsmath}
\usepackage{graphics}
\usepackage{graphicx}
\usepackage{xspace}

\renewcommand*{\k}[0]{\ensuremath{k}\xspace}
\newcommand{\ktree}{\ensuremath{\k^2}-tree\xspace}

\newcommand{\ktriples}{\ensuremath{\k^2}-triples\xspace}
\newcommand{\ktriplesplus}{\ensuremath{\k^2}-triples\ensuremath{^+}\xspace}
\newcommand{\rdfcsa}{RDFCSA\xspace}
\newcommand{\ours}{BMatrix\xspace}

\newcommand{\ktrees}{\ensuremath{\k^2}-trees\xspace}

\newcommand{\kk}{\ensuremath{\k^2}\xspace}

\begin{document}

\title
{\large
\textbf{Revisiting compact RDF stores based on \ktrees
\thanks{Funded by: MICINN-AEI/FEDER-UE RTI2018-098309-B-C32, Xunta de Galicia/GAIN IN848D-2017-2350417;
MICINN-AEI/FEDER-UE RTC-2017-5908-7; Xunta de Galicia/FEDER-UE ED431C
2017/58, ED431G/01, IN852A 2018/14; MINECO-AEI/FEDER-UE TIN2016-78011-C4-1-R,
TIN2016-77158-C4-3-R, TIN2015-69951-R; EU H2020 MSCA RISE BIRDS: 690941 } } }

\author{%
Nieves R. Brisaboa$^{\ast}$, Ana Cerdeira-Pena$^{\ast}$, Guillermo de Bernardo$^{\ast}$ and Antonio Fari\~na$^{\ast}$\\[0.5em]
{\small\begin{minipage}{\linewidth}\begin{center}
\begin{tabular}{ccc}
$^{\ast}$Universidade da Coru\~na, Centro de investigaci\'on CITIC, Databases Lab. \\
%Campus de Elvi\~na s/n  &&\\
%A Coru\~na, 15071, Spain &&\\
A Coru\~na, Spain &&\\
\url{{brisaboa, acerdeira, gdebernardo, fari}@udc.es} &&\\
\end{tabular}
\end{center}\end{minipage}}
}

\maketitle
\thispagestyle{empty}

\begin{abstract}
We present a new compact representation to efficiently store and query large RDF
datasets in main memory. Our proposal, called \ours, is based on the \ktree, a
data structure devised to represent binary matrices in a compressed way, and
aims at improving the results of previous state-of-the-art alternatives,
especially in datasets with a relatively large number of predicates. We
introduce our technique, together with some improvements on the basic \ktree
that can be applied to our solution in order to boost compression. Experimental
results in the flagship RDF dataset DBPedia show that our proposal achieves
better compression than existing alternatives, while yielding competitive query
times, particularly in the most frequent triple patterns and in queries with
unbound predicate, in which we outperform existing solutions.
\end{abstract} 

\Section{Introduction}

The amount of valuable resources publicly available on the Web, in recent years,
has increased to such an extent that new problems have arisen related to processing those resources.
Getting insight into data and extracting knowledge from huge repositories of
information has become a critical task. Based on the principles of the Semantic
Web~\cite{bernerslee2001semantic}, the Web of Data has emerged as an effort to provide 
an environment of common access to the published data, by representing it through standard
formats, so that it can be automatically reachable, and discovered. 

The \textit{Resource Description Framework} 
(RDF)~\cite{Manola2004} is a W3C recommendation to describe any resource in the form of triples \textit{(subject, predicate, object)}.
The popularity of RDF has led to the development of RDF
stores, systems devoted to the storage of RDF data that also provide query
support to access the stored information. The standard language to perform
queries on RDF datasets is SPARQL~\cite{SPARQL:2008}, and basic graph patterns
or \emph{triple patterns} constitute its core. A triple pattern is a tuple $(s,p,o)$, $s \in S$, $p \in P$ and $o \in O$, 
where each element can be set to a value or left unbound. For instance, $(s,p,?)$ matches
all the RDF triples that have subject $s$ and predicate $p$. 

RDF does not enforce an underlying storage format. Hence, a large number of
works have been proposed in the last years to store and query RDF data, ranging from 
relational solutions~\cite{Sakr:2010} to native approaches~\cite{Neumann:2010,MONET,Cureetal.ESWC:2014}. 
As the popularity of RDF has been increasing, so has the size of RDF
repositories. To handle these larger datasets, new approaches have been
proposed: distributed stores~\cite{hadooprdf,dream}, %Quitar hadoop??
and solutions based on compact data structures. For instance,
\ktriples~\cite{sandra2015} relies on vertical partitioning combined with a
compact representation of binary matrices, called \ktree~\cite{ktree}; 
\rdfcsa~\cite{rdfcsa} is based on compressed suffix arrays \cite{Sad03}.

In this paper we introduce a new representation based on \ktrees, called \ours.
Instead of resorting to vertical partitioning like \ktriples, we aim at
storing triples in a few data structures, in order to improve performance in RDF collections with a larger number of predicates.
Particularly, \ours consists of two binary matrices, one related to
triples subjects and the other to objects, and a small additional data
structure; each matrix is stored using a \ktree. Experimental results show that our proposal
beats \ktriples, the most compressed representation in the
state of the art up to now, in terms of space. We are also very
competitive in query times, especially in the most used queries.
Particularly, we are very efficient in queries with unbound predicate, where the
\ktriples needs additional indexes to be competitive in query times.

\Section{Related Work}\label{relatedWork}

In this section we present the basic data structures that are used in the
\ktree, as well as the \ktree itself, since they are necessary for understanding our proposal. 

%\SubSection{Basic concepts}

A \textit{bit sequence} or \textit{bitmap} is a sequence of \textit{n}
bits, $B[1,\textit{n}]$, that supports three basic operations: 
$rank_c(B,i)$ counts the number of occurrences of bit $c$ in $B$ up to position $i$;
$select_c(B,j)$ returns the position in $B$ of the $j$-th bit set to
$c$; and $access(B,i)$ gets the bit value at $B[i]$. These operations 
can be answered in constant time using $o(n)$ bits in addition to the bitmap~\cite{cds}. 
Compressed representations~\cite{sdarray} can answer the same operations while compressing the bitmap.  
In this work we use a practical implementation~\cite{GGMNwea05} that is based on
single-level sampling. The default setup adds a $5\%$ of space overhead and provides efficient query times.

\textit{Directly addressable codes}~\cite{dacs}, or DACs, is a technique that provides
direct access to sequences of variable-length codes, where each codeword can be regarded as a sequence of
chunks of $b$ bits each, for any fixed $b$. DACs works by reorganizing these
chunks in several arrays, $L_i$, and using additional bitmaps, $B_i$, to mark for each entry in $L_i$ 
whether the corresponding word has a next chunk in $L_{i+1}$ or not. In that way, entries can be decompressed 
by accessing the first chunk directly and then using $rank_1$ operations on the $B_i$s to locate 
the corresponding position of the next chunk.

%\SubSection{The \ktree and its application to RDF}

\begin{figure*}[h]
  \includegraphics[width=\textwidth]{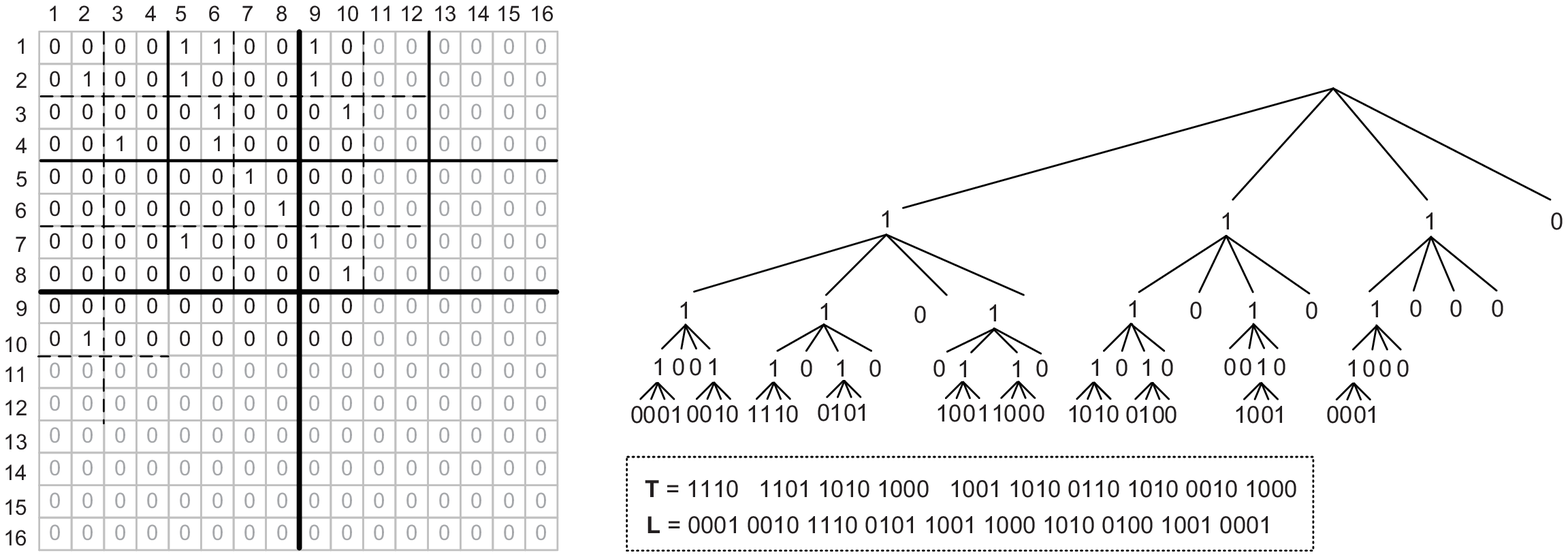}
  \caption{Example of binary matrix and its associated \ktree.}
  \label{fig:k2tree}
\end{figure*}

The \textit{\ktree}~\cite{ktree} is a compact representation of sparse binary
matrices originally devised for Web graphs. Given an $n \times n$ matrix,
it is represented as a $\kk$-ary tree, for a fixed $k$. The root of the tree
represents the complete matrix. The matrix is subdivided in $\kk$ submatrices of
equal size. These submatrices are read in a left-to-right and top-to-bottom
order, and for each of them a child node is appended to the tree root. Each node is marked with a
single bit: 1 if the submatrix contains at least a 1, and
0 otherwise. The decomposition process continues recursively for each 1-bit,
until we reach the cells of the original matrix. The conceptual \ktree is
traversed levelwise, and its bits are stored using just two bitmaps: $T$ stores
the bits from all the levels except the last one; $L$ stores the bits from the last level.
Figure \ref{fig:k2tree} shows a conceptual matrix, its \ktree for
$k=2$, and the corresponding $T$ and $L$ bitmaps.

Single cell, row/column, and bi-dimensional range queries can be answered by means of
traversals of the conceptual tree, starting at the root and traversing only
the submatrices intersecting the queried region.
Top-down traversal of the tree can be replicated in the bitmaps $T$ and $L$
using  the following property: given an internal node at position $p$
($T[p]=1$), its $\kk$ children are consecutive and start at 
position $p'=rank_1(T, p) \times \kk$ in $T$:$L$ (i.e. the concatenation of $T$ and $L$).

Some modifications to the basic \ktree have been presented by the original
authors. The most relevant one is the statistical compression of the lower levels of the conceptual tree to exploit small-scale regularities in 
the binary matrix. This is achieved by building a matrix vocabulary and representing each matrix by its identifier; the sequence of encoded matrices is then
stored using DACs.

The \ktriples~\cite{sandra2015} is a solution based on \ktrees to represent RDF datasets. 
It applies vertical partitioning to the RDF dataset, creating $|P|$ binary
matrices that correspond to the pairs $(s,o)$ associated with each predicate.
Each of those matrices is represented with a \ktree. In this solution, triple
patterns can be easily translated into \ktree operations.
For instance, an $(s,p,o)$ query is solved by checking cell $(s,o)$ in the \ktree associated to $p$. The
query $(s,?,?)$ is translated into $|P|$ queries asking for all the elements in row $s$ in each \ktree. 
Notice that, whenever the predicate of a query is unbound, the \ktriples must
query all the \ktrees, which may be costly in datasets
with a large number of predicates. An enhanced variant of the data structure,
called \ktriplesplus, uses additional indexes to cope with this problem.

\begin{figure*}[h]
  \includegraphics[width=\textwidth]{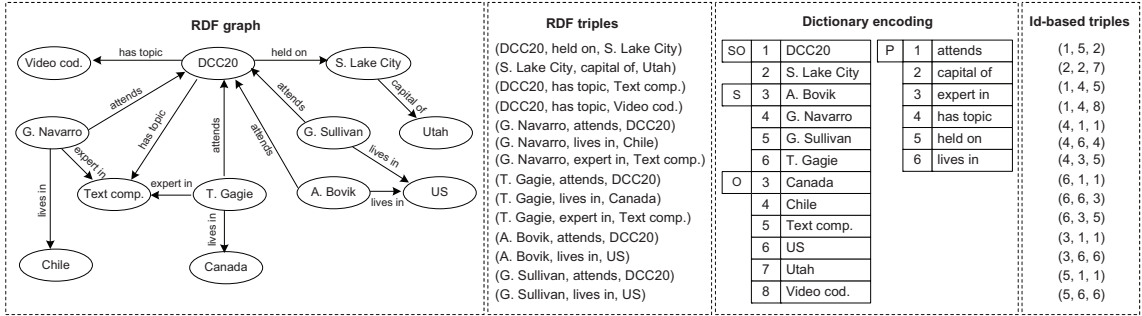}
  \caption{Example of RDF graph and its corresponding dictionary encoding.}
  \label{fig:rdf}
\end{figure*}

\Section{Our proposal: \ours}\label{proposal}

As explained in previous sections, techniques based on \ktrees have shown good
compression capabilities over RDF data. However, the vertical partitioning used in \ktriples makes the structure slow to answer queries with
unbound predicates. Using additional indexes partially solves the problem, but
leads to significantly larger space requirements. Our proposal, that we call \ours, 
also uses the \ktree as underlying data
structure, but we follow a different approach for organizing the RDF data. 
Particularly, our goal is to use fewer structures, and keep all the predicates
together.

Our representation is designed to store RDF triples encoded as integer
identifiers. Therefore, it requires a dictionary to encode/decode the original
strings into integer ids. We follow the same scheme for dictionary encoding 
used by state-of-the art solutions based on compact data
structures like \ktriples and \rdfcsa. Figure~\ref{fig:rdf} shows an
example of RDF graph and the corresponding dictionary encoding. Strings are
divided in four categories: subject--objects (strings that appear as both
subjects and objects), subjects-only, objects-only and predicates. Consecutive
ids are assigned to each unique string in each category (notice that
subject-only and object-only entries start numbering after the last
subject--object).

\begin{figure*}[h]
  \centering
  \includegraphics[width=0.80\textwidth]{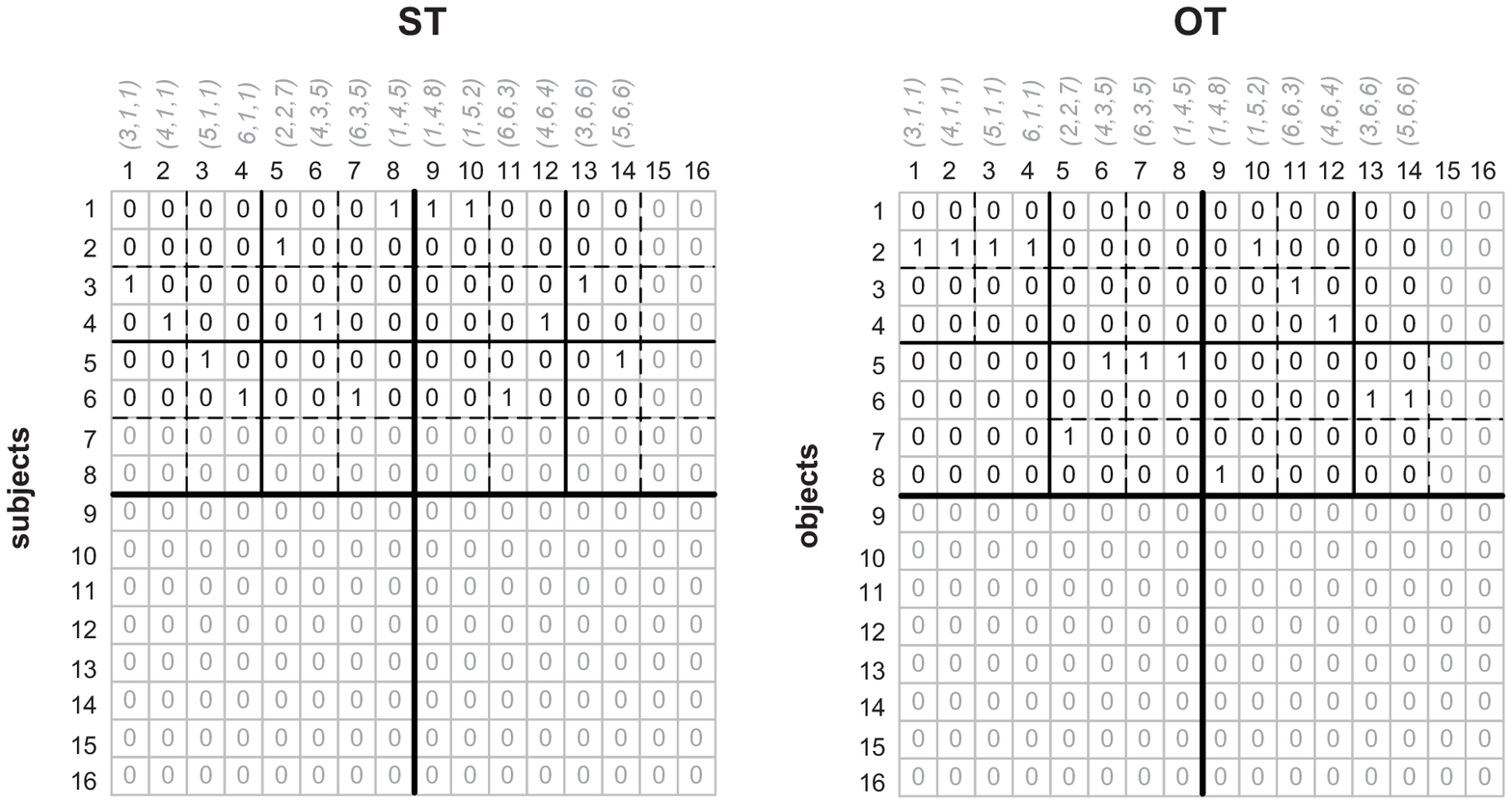}
  \caption{\ours conceptual representation.}
  \label{fig:matrices}
\end{figure*}

After dictionary encoding, we have a collection $T$ of $n$ triples
$t_i=(s_i,p_i,o_i)$, where each $s_i$, $p_i$ and $o_i$ is an integer. 
triples with the same predicate are grouped together. We use
$p,o,s$ order since it leads to better
compression in practice, but any other ordering that
groups triples with the same $p$ could be used. After sorting, we build two
binary matrices $ST$ and $OT$. $ST$ has $|S|$ rows and $n=|T|$
columns, and a cell $(r,c)$ in $ST$ is set to 1 iff $s_c=r$. $OT$
is similar, but has $|O|$ rows, and a cell $(r,c)$ in $OT$ is set to 1 iff
$o_c=r$. Notice that only a single 1 can appear in each column of $ST$ and
$OT$. Figure~\ref{fig:matrices} shows the matrices generated for the RDF dataset
of Figure~\ref{fig:rdf}. Note that the grayed out portions of the matrices do
not belong to the conceptual representation. However, each matrix will then be
stored using a \ktree, that conceptually expands the matrix to the next power of $k$.

In order to recover the original triples, we also need to store an auxiliary
structure to know the column ranges corresponding to each predicate. We can
use any bitmap $BP$ of length $n$, storing a 1 for the positions where the
predicate changes; in this representation, the predicate of a triple $t_i$ can
be computed as $rank_1(BP, i)$, and the starting position of a predicate $p_i$ as $select_1(BP,i)$.
We use a custom representation supporting those operations as follows: we set an array $AP$ of size $|P|$ storing the initial
position of each predicate. Additionally, we select a sampling period $d$
and build an array $rankP$, that stores the predicate that contains triple $d \cdot i$
for $i \in [1..n/d]$. In this representation, $select_1(BP,i)$ is computed as
$AP[i]$. To answer $rank_1(BP,i)$, we use $rankP$ to identify the range of
predicates that could contain that triple $[rankP[i/d], rankP[i/d+1]]$, and
binary search in $AP$ for the rightmost entry that is not greater than $i$.
Assuming that $|P| << n$ and for relatively large $d$, the space required by
this structure is much smaller than $n$.

\SubSection{Query operations}

In this section we describe the implementation of triple pattern queries in
our representation. Essentially, we reduce triple patterns to operations on
\ktrees $ST$ and $OT$ and $rank$/$select$ operations on bitmap $BP$.

$(s,p,o)$ queries just require checking that the triple pattern exists in the
collection. First we find all the triples that have subject $s$ and predicate
$p$. To do this, we compute the range of columns corresponding to $p$ as
$[select_1(BP, p), select_1(BP,p+1)-1]$. Then, we search for all the ones in
$SP$ in row $s$ and in the given range of columns (this operation is implemented
in a \ktree like a simple row search by adding filters at each step that
restrict search to branches inside the column range). For each result $t_i$
found in $ST$, we perform a cell retrieval query in $OT$ for cell $(o,i)$, in
order to check if the triple had $o$ as its object. We return immediately when
a single result is found. Notice that we could also perform the operation
starting in $OT$ and checking in $ST$, but our results suggest that this
alternative is slower due to the large number of intermediate results generated in queries with very common objects.

$(s,p,?)$ queries start, like the previous ones, by finding in $ST$ all the
triples with subject $s$ and predicate $p$. Then, for each result obtained in
$ST$, we need to perform a column query in $OT$ to
obtain the object for that triple. Notice that, since our matrices have a
single 1 per column, column queries can return immediately when they find a
single result. $(?,p,o)$ queries are symmetrical to $(s,p,?)$, starting queries
in $OT$ and then extracting results in $ST$.

$(s,?,o)$ queries are implemented  by first finding all the triples for
object $o$, with a row query in $OT$. Then, we continue
depending on the number of partial results: if the number of intermediate
results is small, we simply check each result $t_i$ in $ST$ with a cell query for $(s,i)$; if the
number of intermediate results is large, we perform a second row query, now in
$ST$, to get all the triples for subject $s$, and intersect both lists to obtain
the final result (the intersection is very efficient since both lists are
already sorted).
In practice, the threshold value $t_{merge-unsorted}$ can be a small value (e.g., 10), since column queries in \ktrees are roughly
an order of magnitude slower than cell retrieval queries.

$(s,?,?)$ queries start again by finding all the triples for subject $s$
with a row query in $ST$. For each partial result $t_i$, we run a column query
in $OT$ to get the corresponding object. Additionally, we must compute the
predicate for each tuple as $rank(BP, i)$. $(?,?,o)$ queries are symmetrical to $(s,?,?)$, performing the row query in $OT$
and the column queries in $ST$.

$(?,p,?)$ queries involve finding all the cells for a given predicate. We start
by obtaining all the subjects for those triples: we compute the column range for
predicate $p$ ($[select_1(BP, p), select_1(BP,p+1)-1]$) and perform a range
query in $ST$ limiting columns to the given range. This yields a list of
$(s,i)$ pairs that will be the results of our query. In order to obtain the
corresponding objects, we again check the number of partial results: if it is
small, we simply perform a column query in $OT$ per result; if it is larger, we
perform a second range query in $OT$, to get a list of $(o,i)$ pairs, and intersect the resulting lists to
  obtain the final result (in this case, lists are not sorted by column, so we
  sort them before merging). We use a different threshold $t_{merge-unsorted}$,
  but again a relatively small threshold can be used in practice
  to guarantee the best overall performance and more stable times in queries
  with many intermediate results.

\SubSection{Space improvements}

Taking advantage of our setup, we can reduce significantly the size of the
vocabulary with simple representations. Consider a matrix vocabulary with $m$ matrices of size $k_L \times k_L$. 
We view the same vocabulary as a set of $mk_L$ columns. We build a bitmap $C$, of
size $mk_L$, so that $C[i]=1$ if the corresponding column has a 1. Then, we use
a separate array $R$ to store the rows containing the ones, requiring $\log_2 k_L$
bits per entry ($k_L$ is assumed to be a power of two).

We can simply create an array $R$ with $mk_L$ entries, and set to 0 columns
without a value. This means that we can store the complete vocabulary using $mk_L(1+\log_2 k_L)$ bits. For any entry $e$ 
  in the vocabulary, we can recover the value at $(r,c)$ in its
  submatrix by checking $C[ek_L+c]$); if it is
  0, the value is 0; if it is 1, we check if $R[ek_L+c]$ is equal to $r$.
  This solution provides minimum query overhead, replacing the bit
  access of the plain vocabulary with a few array accesses and checks.

A more elaborate scheme can be built storing an entry in $R$ only for
  columns that have a 1. In this variant, to
  access position $(r,c)$ in matrix $e$, we first check the
  bit $C[ek_L+c]$, like in the previous alternative; if it is 0, the value is 0; if it is 1, we
  must access $R$ at position $p'=rank_1(C, ek_L+c)$; and check
  if $R[p']$ is equal to $r$. This variant can save a significant amount
  of space compared to the previous one, but has a significant overhead in
  practice due to the complexity of the rank operation.

\Section{Experimental Evaluation}\label{experiments}

Our proposal is designed to work well in datasets with a relatively large number
of predicates, where other alternatives like \ktriples require additional space to
efficiently answer queries. We evaluated the compression and query performance of
our solution using the {\em DBPedia}
dataset\footnote{\texttt{http://downloads.dbpedia.org/3.5.1/}}, a widely used, large and heterogeneous RDF collection. The original
size of DBPedia, considering triples storing string values, is around 34 GB, and
it contains 232M triples. After applying dictionary compression to the
dataset, the collection of triple identifiers can be stored in 2.6GB (i.e., 3
integers per triple). The dataset contains 18.4M different subjects, 39,672
predicates and 65.2M different objects. We use an existing testbed
\footnote{Available at \texttt{http://dataweb.infor.uva.es/queries-k2triples.tgz}, provided by the authors of \ktriples.}
that includes 500 queries of each triple pattern. For each pattern,
we determine a minimum number of repetitions necessary to obtain consistent times and measure the average query times per result.

We compare our representation with two state-of-the-art approaches based on
compact data structures: \ktriples and its extension \ktriplesplus, and the \rdfcsa. Both
techniques have been shown to overcome alternative solutions in space, and
provide very efficient query times for most triple pattern queries. For \rdfcsa
we use the default configuration, and test sampling values  $t_{\Psi} \in \{16,32,64,256\}$; among the query
implementations provided by the authors, we show results in our experiments for
the binary search, that can be applied to all triple patterns and is the most
consistent in query times. For \ktriples, we use a hybrid representation with $k=4$ in the first 5 levels
of decomposition and $k=2$ in the remaining levels. The bitmaps use the default rank structure, requiring an extra
5\% space. The lower levels of the tree are compressed using a matrix vocabulary
of $8 \times 8$. In queries that have unbound predicate, we show the tradeoff
obtained by the basic version (smaller, slower) and the \ktriplesplus version
with additional indexes (larger and faster).

To provide the fairest possible comparison with \ktriples, our \ktrees use the
same exact configuration, the only difference being the vocabulary
representation: we use the simplest representation of the vocabulary proposed in
the previous section, in order to obtain some space savings with minimal
overhead. Additionally, we show results for our representation with denser
sampling in the \ktrees bitmaps, requiring 12.5\% extra space; this leads to a larger solution
with faster query times. This is used just to outline the level of tradeoff that
can be obtained tuning this parameter in \ktrees; notice that a similar tradeoff
can be achieved in \ktriples.

We run our experiments on an Intel Xeon E5-2470@2.3GHz (8 cores)
CPU, with 64GB RAM. The operating system
was Debian 9.8 (kernel 4.9.0-8-amd64). Our code is implemented in C and compiled
with gcc 6.3.0 with the -O9 optimization flag.

\SubSection{Results}

\begin{figure}[t!]
    \centering
    \includegraphics[angle=-90,width=0.34\textwidth]{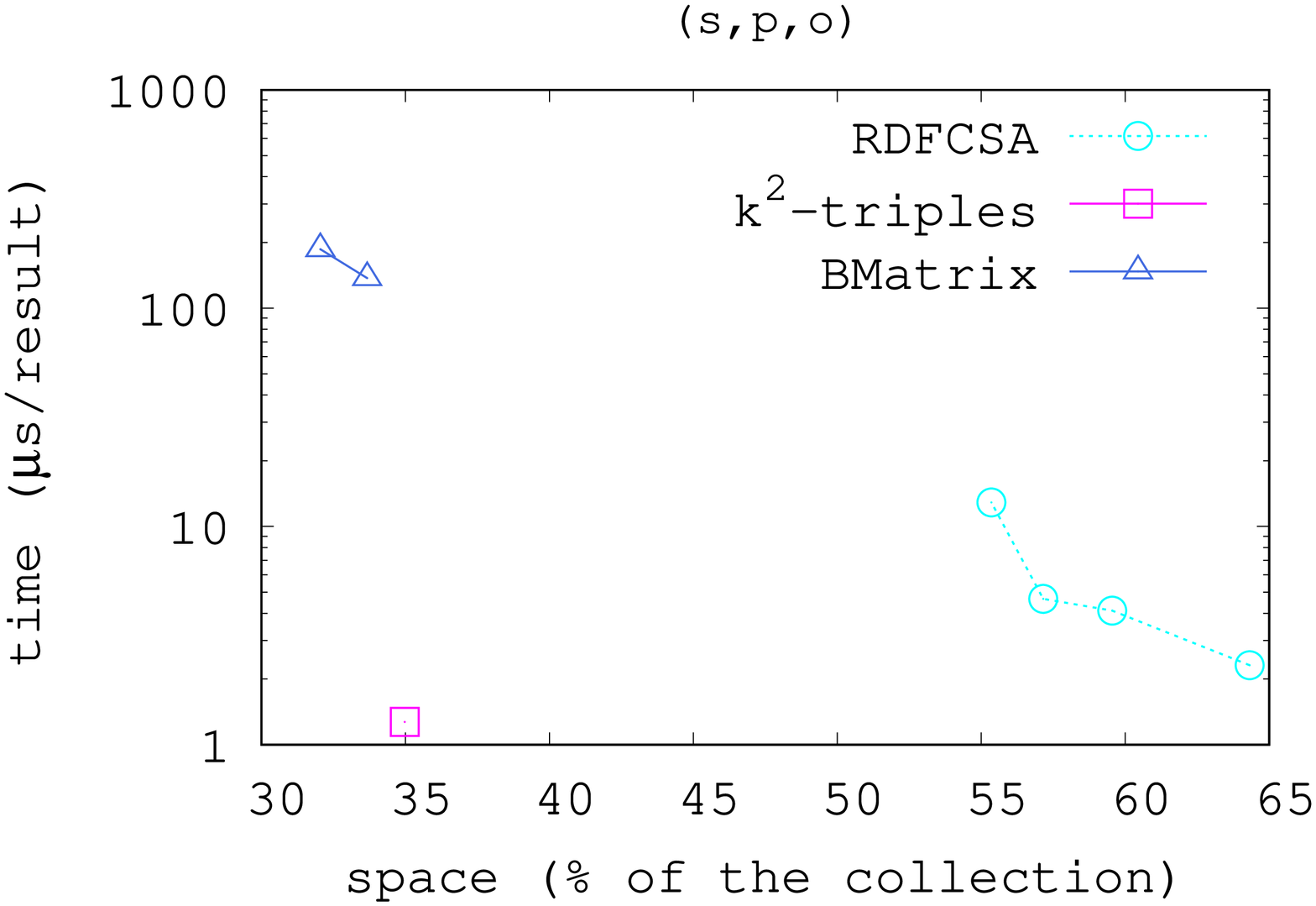}
    \includegraphics[angle=-90,width=0.34\textwidth]{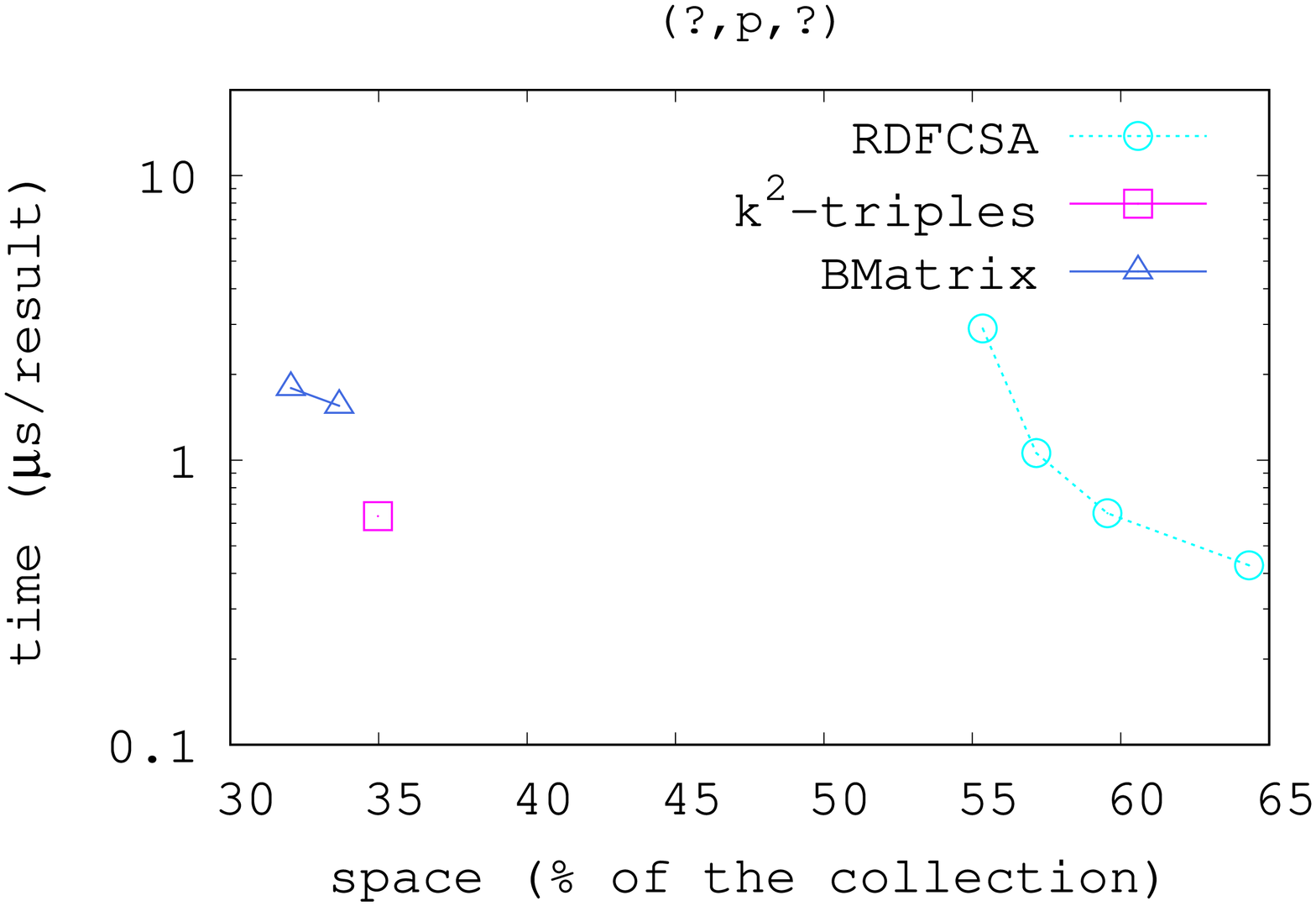}
    \\
    \includegraphics[angle=-90,width=0.34\textwidth]{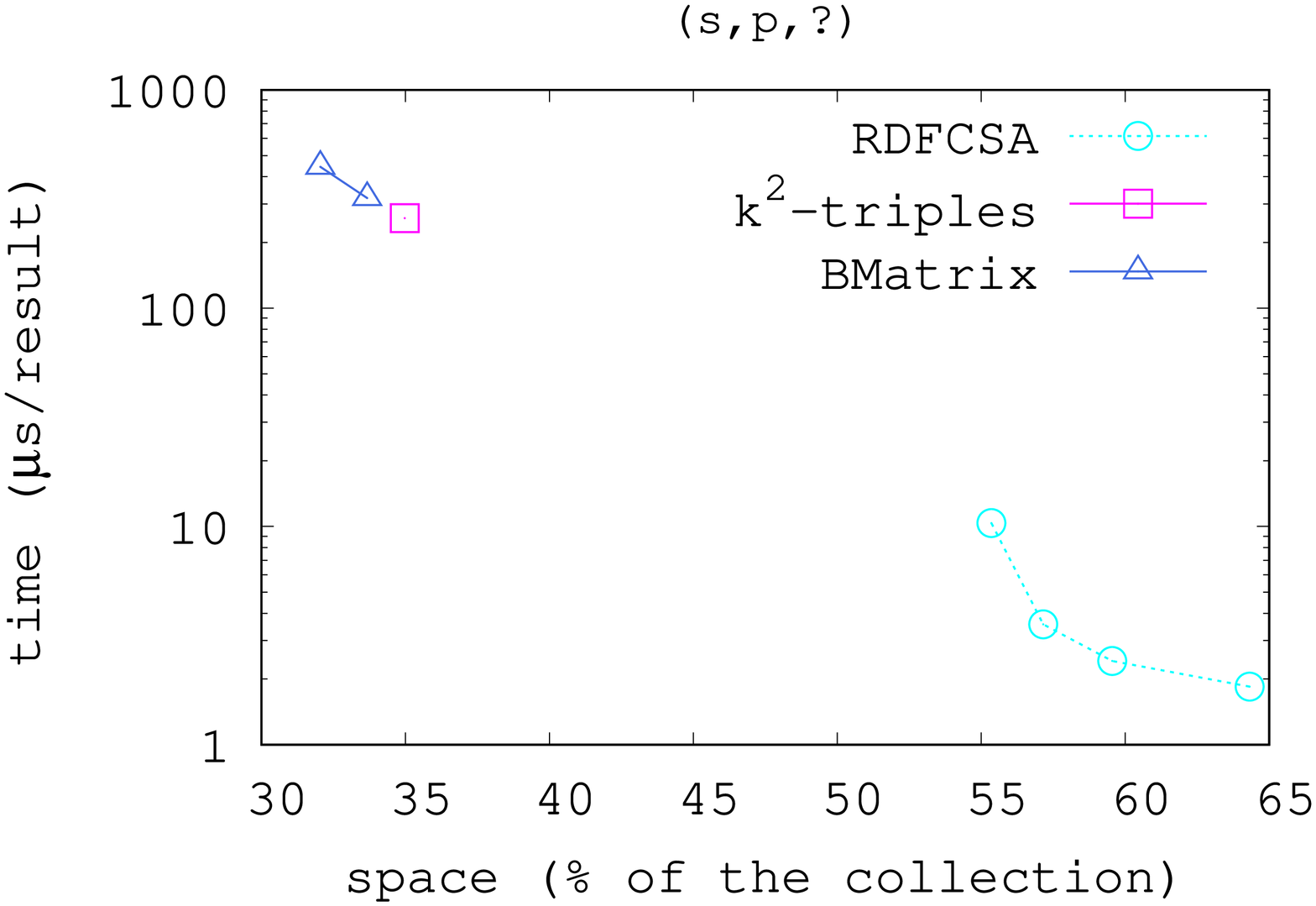}
    \includegraphics[angle=-90,width=0.34\textwidth]{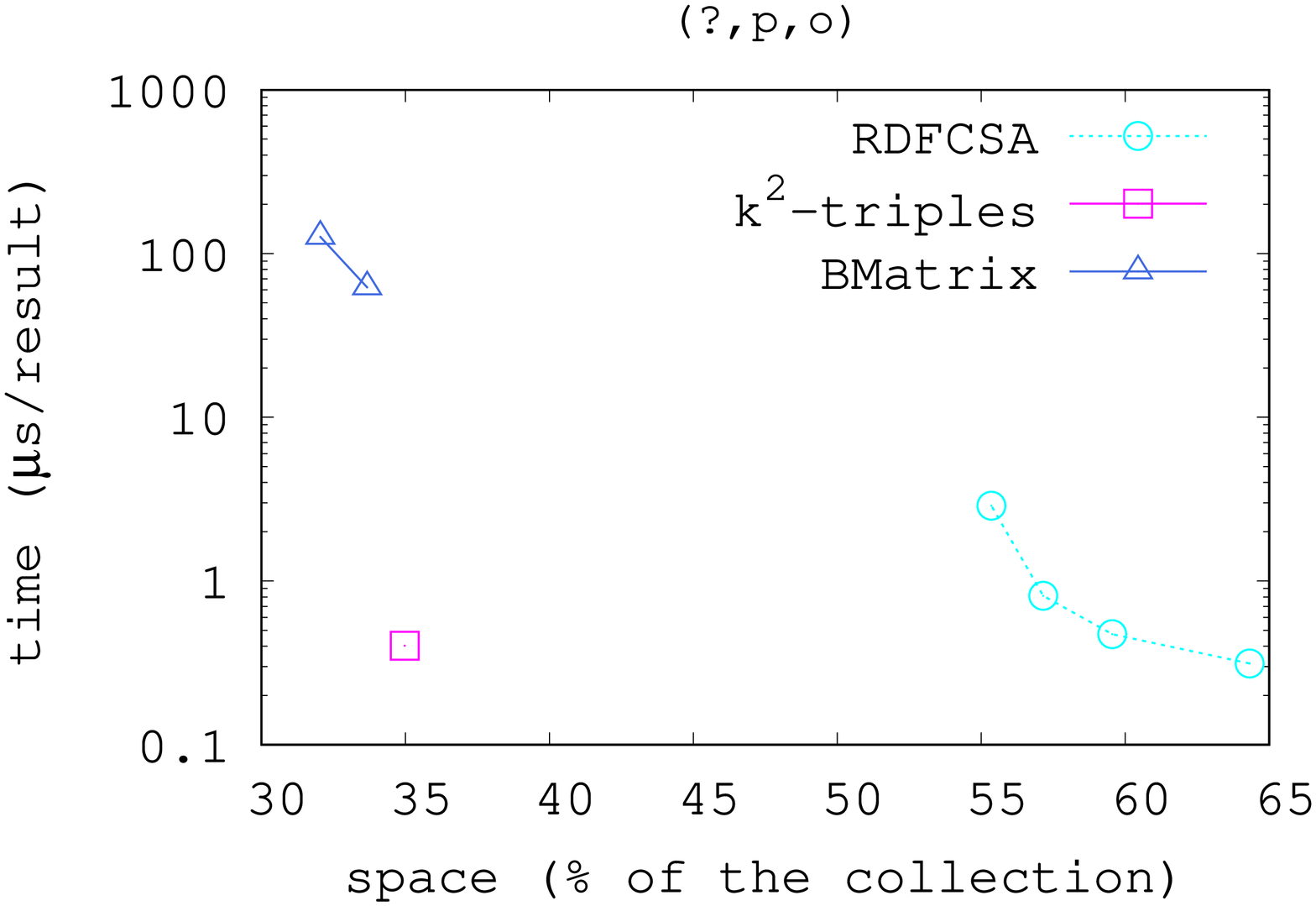} \\
    \includegraphics[angle=-90,width=0.34\textwidth]{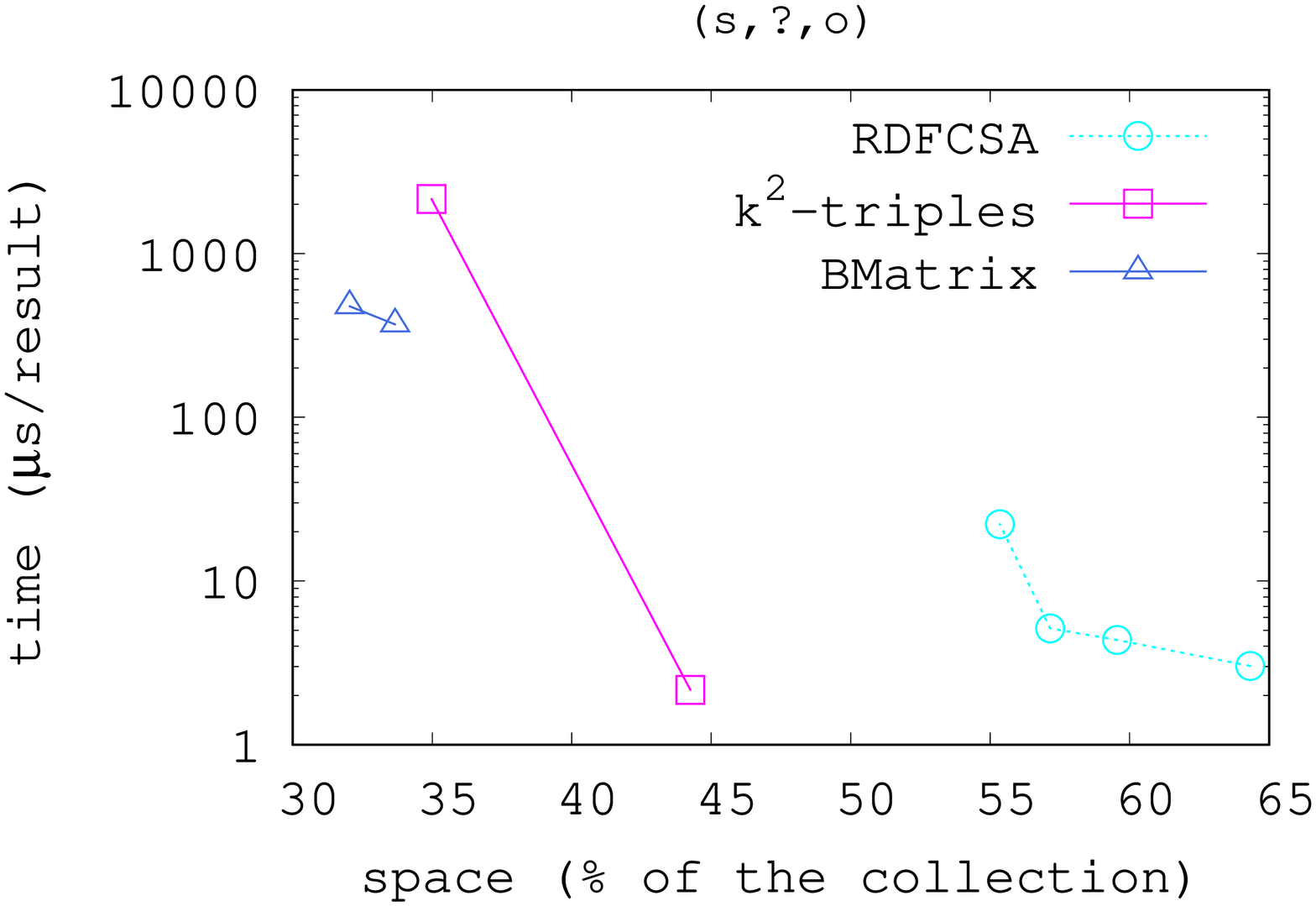}\hspace{-0.5cm}
    \includegraphics[angle=-90,width=0.34\textwidth]{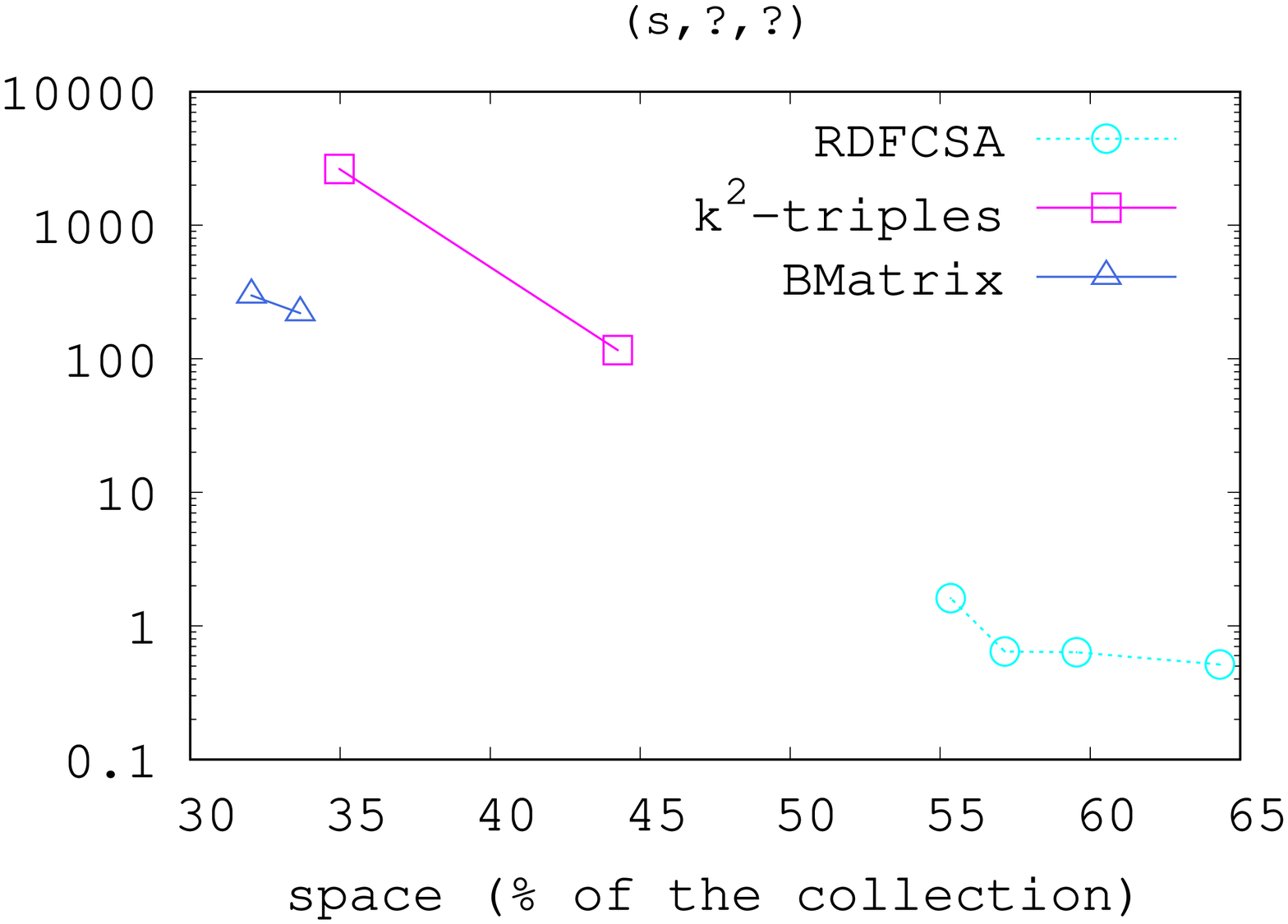}\hspace{-0.5cm}
    \includegraphics[angle=-90,width=0.34\textwidth]{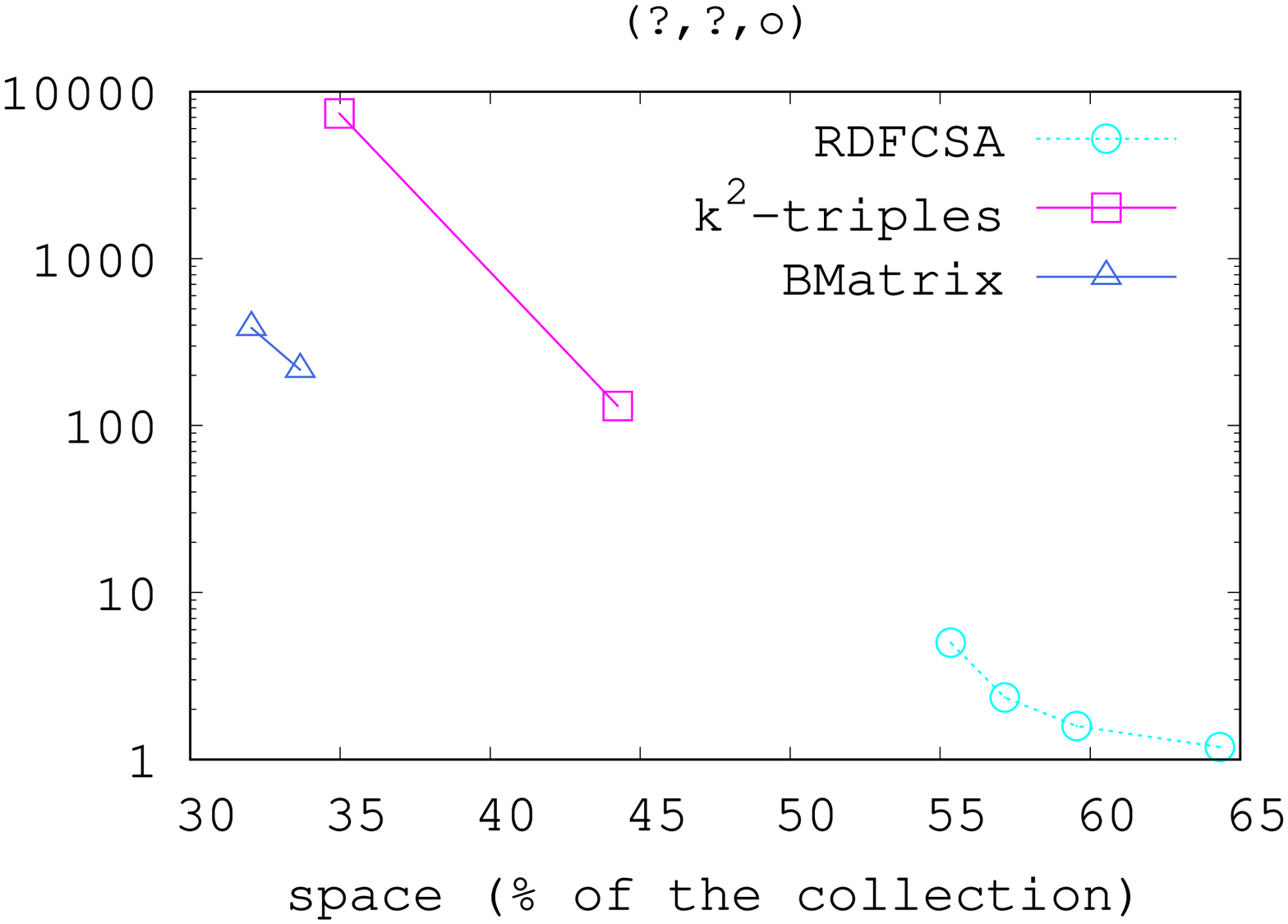}
    \caption{\label{fig:queriesBP} Space/time tradeoff for triple patterns. Space
  measured as percentage of the input size. Query times in $\mu$s per
  result.}
\end{figure}

We measure the compression and query efficiency of our proposal on the seven
triple patterns that compose the basis of SPARQL queries. We divide our
experimentation in two main groups of patterns: the four plots at the top of
Figure~\ref{fig:queriesBP} display results for triple patterns with fixed predicate,
while the three at the bottom display results for patterns with unbound
predicate.

Results show that \ours significantly improves the
compression of \ktriples. With sampling similar to \ktrees, we require
10\% less space. Even using the denser sampling, we are still
smaller than \ktriples. Furthermore, when \ktriples needs to use additional
indexes it becomes much larger than \ours. \rdfcsa is significantly larger
than any of the other alternatives.

As shown in Figure~\ref{fig:queriesBP} (top), query times for patterns with fixed predicate in \ours are
quite consistent, but comparison results are significantly different depending on the triple pattern:
in $(s,p,o)$ queries, \ktriples just needs to run a cell query in a \ktree,
whereas our algorithm is more complex; this leads to our implementation becoming
significantly slower. In $(?,p,?)$ queries, we are still slower than \ktriples
but much closer, since range operations are required in both cases; query times
are also comparable to those of \rdfcsa using about half their space. In
$(s,p,?)$ queries we are comparable to \ktriples, but \rdfcsa is very
efficient and provides an interesting tradeoff.
In $(?,p,o)$ queries \ours is again significantly slower than \ktriples, that is
clearly the best alternative. Notice that this query is essentially equivalent
to the previous one in complexity, both in \ktriples and \ours, but \ktriples is
much more efficient in $(?,p,o)$, due to the usually much larger number of results
per query in $(?,p,o)$ queries. In \ours, we
have to extract the object individually for each result, so having a few queries that yield many results
has a significant impact on our performance.

% \begin{figure*}[htbp]
% \centering
%   \includegraphics[angle=-90,width=225pt]{data/so}\\
%   \includegraphics[angle=-90,width=225pt]{data/s}
%   \includegraphics[angle=-90,width=225pt]{data/o}
%   \caption{Space/time tradeoff for triple patterns with unbound predicate.
%   Space is measured as percentage of the input size. Query times are $\mu$s per
%   result.}
%   \label{fig:queriesUP}
% \centering
% \end{figure*}

For triple patterns with unbound predicate,
Figure~\ref{fig:queriesBP} (bottom) displays two points as space/time tradeoff
for \ktriples; these correspond to the basic implementation and
the \ktriplesplus variant with additional indexes. Results show that \ours
significantly improves the query times of the basic \ktriples implementation. In
$(s,?,o)$ queries, \ktriples is still the fastest technique when using extra
indexes, but to do this it requires 40\% more space than \ours. In $(s,?,?)$ and
$(?,?,o)$ queries, we are an order of magnitude faster than \ktriples without
extra indexes, and even comparable in query times to the \ktriples version that
uses extra 40\% space. \rdfcsa is faster than \ours, but almost twice as large,
so our proposal is still the best option when memory usage is an issue.

Taking into account the different comparison results obtained, \ours provides a very reasonable space/time tradeoff depending
on the types of queries to be executed: when a large percentage of triple
patterns with unbound predicate are expected, \ours clearly overcomes
\ktriples and provides a very compact alternative to \rdfcsa. Additionally, our
experiments show that \ours is more competitive in the triple patterns that are
more frequently used: a previous analysis on the DBPedia dataset~\cite{AFMPF11}
has shown that 90\% of the triple patterns used in SPARQL
queries over DBPedia are $(s,p,?)$, where \ours is competitive with \ktriples,
and $(s,?,?)$, where \ours is either much faster or much smaller than the
alternatives.

\Section{Conclusions and Future Work}\label{conclusions}

We have introduced \ours, a compact representation of RDF datasets based on \ktrees. It
aims mainly at improving the performance of previous solutions in datasets with a 
relatively large number of predicates, where the vertical partitioning strategy 
leads to poor query times in patterns with unbound predicate. As a side
result, we also propose some simple improvements on the \ktree data
structure that have stand-alone interest and could be applied to other domains.

We experimentally evaluate our proposal on DBPedia, a widely used
RDF dataset containing around 40,000 predicates. We compare
our proposal with \ktriples and \rdfcsa.
Our results show that \ours achieves better compression,
being 10\% smaller than \ktriples and 40--50\% smaller than
\rdfcsa. \ours is also competitive in query times in the most frequent query patterns. 
In query patterns with unbound predicate, \ours is faster than the basic
\ktriples. For $(s,?,?)$ and $(?,?,o)$ queries, we obtain query times comparable to those of the most efficient \ktriples
version with extra indexes, that uses 40\% more space than our proposal.

Currently, \ours supports all basic triple patterns. We plan to
extend our evaluation to multi-pattern join queries, that can be supported
by merging or chained evaluation of individual triple patterns, as in
state-of-the-art alternatives. Synchronized traversal of multiple \ktrees, used
in \ktriples, can also be applied to our solution in order to improve query
times. Finally, we believe that new tradeoffs can be obtained in
solutions based on \ktrees to speed up specific queries. \ours aims at
boosting queries with unbound predicate, that are relevant for many
application domains, but some other arrangements could benefit other application scenarios.

% Different arrangements of the input triples, and improvements on the underlying data
% structures, could lead to even better compression. For instance, \ours sorts the
% triples in $p,o,s$ order, which yields a relatively small $OT$ tree and a
% larger $ST$; triple arrangements that work for both trees could significantly
% reduce the space of our proposal.

\Section{References}
\bibliographystyle{IEEEbib}
\bibliography{refs}

\end{document}